\documentstyle[10pt,epsfig]{article}
\newenvironment{mytitle}{\begin{center} \LARGE}{\\ [.1in]\end{center}} 
\newenvironment{myauthor}{\begin{center} }{\\ [.1in]\end{center}} 
\newenvironment{myinstit}{\begin{center} \it}{\end{center}}

\begin{document}

\begin{mytitle}
From Chaos to Disorder in Quasi-1D Billiards with Corrugated Surfaces
\end{mytitle}
\vspace{.5cm}

\begin{myauthor}
J. A. M\'endez-Berm\'udez, G. A. Luna-Acosta, and F. M. Izrailev
\end{myauthor}

\begin{myinstit}
Instituto de F\'{\i}sica, Universidad Aut\'onoma de Puebla, Apartado Postal 
J-48, Puebla 72570, M\'exico\\
\end{myinstit}

\begin{abstract}
We study chaotic properties of eigenstates depending on the degree of complexity in boundaries of a 2D periodic billiard. Main attention is paid to the situation when the motion of a classical particle is strongly chaotic. Our approach allows to explore the transition from deterministic to disordered chaos, and to link chaos to the degree of ergodicity in eigenstates of the billiard. We have found
that bouncing balls strongly reduce chaotic properties of eigenstates, thus leading to a serious problem in statistical description for global properties of eigenstates. A quite unexpected effect of rough surfaces on the form of eigenstates has been discovered and explained by a strong localization of a subset of eigenstates in the energy representation.
\\


\end{abstract}

\vspace{.5cm}

\section{Introduction}

Electromagnetic, acoustic, or quantum waves are essential means for communications and measurements and hence their propagation and scattering properties are of great scientific as well as technological interest. During the last decade or so, there has been a surge of theoretical and experimental breakthroughs inspired mostly by the phenomenon of localization in disordered media \cite{sheng,boris}. Localization is a wave interference effect and hence
there is a host of potentially rich phenomena including, weak localization of light, Anderson localization and transport of electromagnetic radiation in random media, localization of surface gravity waves on a random bottom, and 
electron/wave propagation in films or waveguides with rough surfaces \cite{furutsu,rossum,janssen}. The well known result from the scaling theory of localization \cite{Abrahams} is that for two dimensional samples of infinite extent, there will  be no propagation of waves and the stationary solutions are exponentially localized if the disorder is strong enough. For one dimension, any amount of disorder is sufficient. More recently there have been studies on correlated random models. In one dimension, the latter allows for the appearance of the mobility edge, envisioned more than a decade ago but just theoretically \cite{kriz} and experimentally \cite{kriz2} established a few years ago. Disorder can be caused by the medium or by the boundaries between media as may be the case for ballistic mesoscopic systems or for gravity waves at the surface
of the water in the oceans. 

Localization is known to occur not only in random models but also in deterministically quantum chaotic systems \cite{locall,I90}. This type of localization is known as dynamical localization since it is entirely due to the deterministic chaotic dynamics, where random or disordered potentials are absent. In the regime of fully chaotic dynamics (hard chaos) some properties of these systems may  be described rather well by the same mathematical methods of disordered systems. In particular, by exploiting the nonlinear sigma model, Efetov \cite{Efetov} has shown that on long time scales such properties as level statistics of weakly disordered systems are universal and coincide with those of Random Matrix Theory (RMT). The latter are known to serve for description of quantum systems that are classically chaotic \cite{I90,linda,stockmann}. Additionally, based on Gutzwiller's formula \cite{gutz}, semiclassical calculations \cite{berry1} predict new specific correlations in energy spectra, similar to what was obtained by another mathematical tool from disordered systems, the diagrammatic perturbation theory \cite{szafer}.

While there are some properties that are shared by disordered and chaotic systems, the common approaches employed are very good insofar as they reveal universal properties under certain conditions. Thus, theories such as RMT or the nonlinear sigma model are unable to describe particular features (or deviations from the universal predictions) that will distinguish between one chaotic system and another. One important advantage of a deterministic approach to quantum chaos is that it gives us the possibility of understanding particular quantum features in terms of its underlying classical dynamics. It is well known, for example, that scars and bouncing ball states \cite{vergini} are responsible for deviations from the predictions of Schnirelman's theory \cite{schni} for wave functions and also from the level statistics of RMT \cite{porter}. Further, in \cite{gala2,gala3,gala4} it is shown that specific features of the structure of wave functions of quantum chaotic system are manifestations of certain type of periodic orbits of the underlying classical dynamics.

In this work we examine the transition from chaos to disorder in a completely deterministic manner. We study the solutions of the Schr\"{o}dinger and Newton equations of a system which is deterministically chaotic in the classical limit, then we examine the transition chaos-disorder by gradually increasing the degree of complexity of the ``interaction'' potential till the scattering potential can be considered practically as a random one. As a paradigm we use a 2D chaotic billiard that is periodic in one direction. Specifically, the billiard is defined by two parallel hard walls, a flat one at $y=0$ and a modulated one given by $y=d+a\xi(x)$, where $\xi$ is periodic $\xi(x+b)=\xi(x)$ in the longitudinal coordinate $x$ with period $b$, $d$ is the average width of the billiard and $a$ is the modulation amplitude. Here $x$, $y$, $d$, $a$, and $b$ are dimensionless quantities obtained from $Y=D+A\xi(2\pi X/B)$ scaled to the period $B$. Then, $x=2\pi X/B$, $y=2\pi Y/B$, $d=2\pi D/B$, $a=2\pi A/B$, and $b=2\pi$, where $X$, $Y$, $D$, $A$, and $B$ have dimensions of length. The case $\xi(x)= \cos(x)$, the so-called ``cosine'' or ``ripple'' billiard \cite{gala2,gala3,gala4,gala0,gala1,gursoy,ketz,jamb1,jamb2}, is known to yield the generic transition to chaos of Hamiltonian systems. Various transport and dynamical properties of the cosine billiard have been investigated in both, classical \cite{gala0,jamb1,jamb2} and quantum regimes \cite{gala2,gala3,gala4,gala1,gursoy,ketz,jamb1,jamb2} in its two versions: finite \cite{gala0,gursoy,ketz,jamb1,jamb2} (one or few periods of modulation) and infinite \cite{gala2,gala3,gala4}. Here we are interested in studying the properties of the eigenstates as a function of the complexity of the top profile. For this purpose we write
\begin{equation}
\xi(x)=\sum_N^{N_T} A_N \cos(Nx),
\label{xi}
\end{equation}
where $A_N$ are random numbers in the range [-1,1]. This form allows us to choose the desire degree of complexity of the top boundary; from smooth ($N_T \sim 1$) to extremely rough ($N_T \approx 100$), see Fig. 1. We remark that even for a single harmonic [$N_T=1$ in (\ref{xi})] the dynamics can yield mixed or full chaos, depending on the values of the parameters $a$, and $d$. Here we consider only the case of full chaos. Note also that $\mid d \xi(x)/dx \mid$ increases in magnitude as $N_T$ increases, unlike the rough profile used in \cite{shepe} where the amplitude of each harmonic $N$ in the sum (\ref{xi}) is inverse proportional to $N$. Therefore, in our case for any given $N_T$ the fluctuations are sharper, more dense, and more ``random'' than that used in \cite{shepe}.

\begin{figure}
\centerline{\epsfig{figure=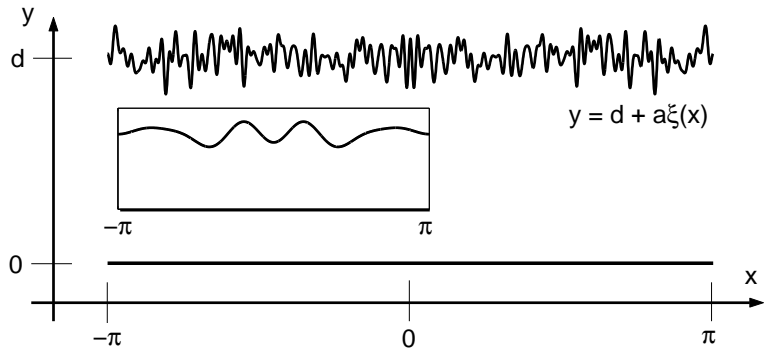,width=4in}}
\caption{Geometry of the modulated billiard for $N_T=100$. The inset shows one period of the billiard for $N_T=6$.}
\label{fig1}
\end{figure}

This paper is organized as follows. In the next section we briefly review the formalism that enables us to incorporate into the Hamiltonian, as an effective potential, all scattering effects that are due to the modulated boundary. There, we also examine the structure of the Hamiltonian matrix, from which we can anticipate the existence of extremely {\it energy-localized} states. In Section 3, we present our calculations of the energy-localization lengths for eigenstates and study them as the complexity of the profile increases. In Section 4, we explore the structure of the {\it energy-localized} states in the case of $N_T= 100$, when the top profile may be considered as random, or rough. In contrast, in Section 5 we discuss the structure of the {\it energy-extended} states as the profiles changes from smooth to rough. In Section 6 we study how the {\it energy-localized} and {\it extended} states manifest themselves in the energy-band structure. Finally, in Section 7 we give some concluding remarks.

\section{Structure of the Hamiltonian Matrix and Bouncing Ball States}

Much information can be extracted from the knowledge of the structure of the Hamiltonian matrix which, of course, depends on the choice of basis used to expand the states. Thus, it is necessary to explain briefly the procedure used in the construction of the Hamiltonian matrix.

\subsection{Coordinate Transformation and Hamiltonian Matrix}

We first perform a transformation to new coordinates, $(x,y)\rightarrow(u,v)$, such that all effects due to the modulated boundary become incorporated into the Hamiltonian operator. This can be accomplished by the curvilinear coordinates $u = x$ and $v =y d/[d+ a \xi(x)]=y/[1+\epsilon \xi(x)]$ ($\epsilon\equiv a/d$). In these coordinates the boundary conditions become $\psi(u,v)=0$ at $v=0$, 1 and the Hamiltonian attains the form $\hat H= (1/2m_e) g^{-1/4}\hat P_\alpha g^{\alpha \beta} g^{1/2} \hat P_\beta g^{-1/4}$ $(\alpha,\beta=u,v)$, where $g^{\alpha \beta}$ is the metric tensor given by 
\begin{equation}
g^{\alpha \beta} = \left( \begin{array}{cc}
1 & {\displaystyle -\frac{v\epsilon \xi_u}{1+\epsilon \xi}} \\
\\
{\displaystyle -\frac{v\epsilon \xi_u}{1+\epsilon \xi}} & 
{\displaystyle \frac{1+(v\epsilon \xi_u)^2}{(1+\epsilon \xi)^2}} 
\end{array} \right).
\label{gab}
\end{equation}
Here $\xi_u \equiv \partial \xi /\partial u$ and $g=(1+\epsilon \xi)^2$ is the metric, see Ref. \cite{gala1} for details. Further, $\hat{H}$ can be splitted into an unperturbed part $\hat{H}^0=(1/2m_e)(\hat{P}_u^2+\hat{P}_v^2)$ and an effective interaction potential $\epsilon \hat{V}(u,v,\hat{P}_u,\hat{P}_v)$.

Since the Hamiltonian is periodic in $u$, the energy eigenstates are Bloch states: $\psi_E(u,v)=\exp(iku)\psi_k(u,v)$, $\psi_k(u,v)=\psi_k(u+b,v)$, where $k$ is the quasi-momentum chosen to lie in the first Brillouin zone $-1/2\leq k \leq 1/2$. The $\alpha$-th eigenstate of energy $E^{\alpha}(k)$ can be expanded as
\begin{equation}
\psi^{\alpha}(u,v;k)= \sum_{m=1}^{\infty}
\sum_{n=-\infty}^{\infty}C_{mn}^{\alpha}(k) \phi_{mn}^k(u,v),
\label{psi}
\end{equation}
where 
\begin{equation}
\phi_{mn}^k(u,v)\,=\,<u,v \mid m,n>_k=\pi^{-1/2}g^{-1/4}\,\,\sin \left( \frac{m \pi v}{d} \right)\,\,
\exp\left[{i(k+n)u}\right]
\label{phi}
\end{equation}
are the eigenstates of $\hat H^0$ with eigenvalues
\begin{equation}
E^0_{n,m}(k) =
\frac{\hbar^2}{2m_e}\left[(k+n)^2+\left(\frac{m\pi}{d}\right)^2\right].
\label{e0}
\end{equation}

The resulting Hamiltonian matrix in the basis of $\hat{H}^0$ is \cite{gala1}
\begin{eqnarray}
H^k_{mn,m'n'} & = & \frac{\hbar^2}{2} (n+k)^2 \delta_{n,n'} \delta_{m,m'} 
+ \frac{\hbar^2}{2\pi} \left( \frac{m^2\pi^2}{2d^2}J^3_{n'n} + 
\frac{\epsilon^2m^2\pi^2}{6}J^4_{n'n} + \frac{\epsilon^2}{8}J^4_{n'n} 
\right) \delta_{m,m'} \nonumber \\
& + & \frac{\hbar^2}{2\pi} \left[ (-\epsilon i)(-1)^{m+m'} \frac{mm'}
{m^2-m'^2} (n+n'+2k)J^5_{n'n} \right. \\
& + & \left. \epsilon^2(-1)^{m+m'} \frac{2mm'(m^2+m'^2)}
{(m^2-m'^2)^2}J^4_{n'n} \right], \nonumber
\label{hmn}
\end{eqnarray}
where
\begin{eqnarray}
J^3_{n'n} & = & \int^{2\pi}_0 du \frac{e^{i(n'-n)u}}{(1+\epsilon\xi)^2}, 
\nonumber \\ 
J^4_{n'n} & = & \int^{2\pi}_0 du \frac{e^{i(n'-n)u}\xi^2_u}{(1+\epsilon\xi)^2}, 
\\ 
J^5_{n'n} & = & \int^{2\pi}_0 du \frac{e^{i(n'-n)u}\xi_u}{(1+\epsilon\xi)}. 
\nonumber
\label{js}
\end{eqnarray}

In a numerical study the Hamiltonian matrix is truncated: $-N_{max}\leq n \leq N_{max}$, $1\leq m \leq M_{max}$. The structure or form of $H^k_{mn,m'n'}=<m',n'|\hat{H}|m,n>$ depends on the ordering of the unperturbed basis $|l>\equiv |m,n>$. That is, we must assign an index $l$, labeling the basis state $|l>_k\equiv |m,n>_k$, to each pair ($m,n$) of indices. The spectra is, of course, independent of the assignment but there are two different assignments which are more convenient for our analysis. To study the structure of the eigenstates, the most physical assignment $(m,n)\rightarrow l$ is, as it will become clear below, such that the unperturbed energy spectrum is ``energy ordered'': $E^0_{l+1}>E^0_l$, $l=1,2,\dots$. On the other hand, to analyze the properties of the Hamiltonian matrix the more convenient assignment is as follows. Starting from the lowest value of $n$ ($n=-N_{max}$) we run through all values of $m$ ($1\leq m \leq M_{max}$). This gives $l=1,2, \dots ,M_{max}$. Then we do the same for $n=-N_{max}+1$, which gives $M_{max}+1\leq l \leq 2M_{max}$, and so on untill we reach $n=N_{max}$ to get $1\leq l \leq L_{max}$, where $L_{max}=(2N_{max}+1)M_{max}$ is the size of the Hamiltonian matrix. This rule results in a block structure of $H_{l,l'}$ containing $(2N_{max}+1)\times (2N_{max}+1)$ blocks, labeled by the pair $(n,n')$ and each of size $M_{max}\times M_{max}$.

\begin{figure}
\begin{center}
{\bf \Large FIGURE 2}\\

{\small (available in gif format)}
\end{center}
\caption{Central part of the $4030 \times 4030$ Hamiltonian matrix
$H_{l,l'}(k)$ with $N_{max}=32$, and $M_{max}=62$. The $62 \times 62$
blocks corresponding to $n,n'=[0,10]$ are shown for the cases (a) $N_T=1$, 
(b) $N_T=2$, (c) $N_T=4$, and (d) $N_T=6$.}
\label{fig2}
\end{figure}

Fig. 2 shows the central part of the $4030\times 4030$ Hamiltonian matrix $[(N_{max},$ $M_{max})=(32,62)]$ for the cases $N_T=1$, 2, 4, and 6 harmonic terms in the profile. A quick look at these figures reveals that the size of the bandwidth increases as $N_T$ increases. By considering the matrix elements $H_{ll'}$ of zero and first order in $\epsilon$ [see Eq. (6)], it can be shown that the band size width $\Delta W$ increases with $N_T$ as
\begin{equation}
\Delta W = (2N_T+1)M_{max}, \quad \quad \quad \epsilon\ll 1.
\label{dw}
\end{equation}

Moreover, the matrix elements with coefficients $\epsilon^2$ give rise to a secondary band that is twice as wide as $\Delta W$, but whose elements lie mainly along the diagonal of the blocks defined by $|n'-n|=2N$, $N=[1,N_T]$. Inspection of Figs. 2(a-d) shows that such estimates agree very well with the numerically computed Hamiltonian matrices. The analysis also shows that the secondary band becomes more important as the number of harmonics $N_T$ increases. This effect is mainly due to the increase, as $N_T$ increases, of the derivative term $\xi_u^2$, appearing in the integrand of $J_{n',n}^4$ [Eq. (7)] which comes with all $\epsilon^2$ terms in $H_{l,l'}$. If the amplitudes $A_N$, defining the profile (\ref{xi}) were inversely proportional to $N$ (as in Ref. \cite{shepe}), then the corresponding Hamiltonian would couple roughly half as many unperturbed states. Thus, the degree of complexity of the boundary profile is manifested in the number of unperturbed states required to construct the eigenstates of the full Hamiltonian $\hat{H}=\hat{H}^0+\epsilon \hat{V}$.

\subsection{Bouncing Ball States}

Another very important dynamical feature, generic of this class of billiards with small amplitude deformations, can be deduced from the structure of $H^k_{mn,m'n'}$ by observing that the matrix elements $<n,m=1|\hat{H}|n',m'=1>\delta_{nn'}$ couple very weakly to the rest of the unperturbed states [see white lines delimiting the blocks in Fig. 2(a-d)]. This fact results in a set of eigenstates of $\hat{H}=\hat{H}^0+\epsilon \hat{V}$ that are practically the same as the eigenstates of $\hat{H}^0$ [see Eq. (4)] (unperturbed states) with $m=1$. We can understand the physical origin of these weakly perturbed states, which we shall refer to as $m=1$ states, as follows. An unperturbed eigenstate (that of a flat channel) has a ``transversal wavelength'' $\lambda_y\equiv 2\pi/k_y=2\pi/(m\pi /D)=2D/m$. Then for $m=1$ states, $\lambda_y$ is so much larger than the amplitude of the ripple ($\lambda_y/a=2/\epsilon\gg 1$) that the state does not resolve the rippleness and hence it is hardly perturbed. On the contrary, unperturbed states with large values of $m$ are strongly affected, and many unperturbed states participate in the formation of such eigenstates of $\hat{H}$. These expectations were numerically tested for the case $N_T=1$ in Refs. \cite{gala2,gala3,gala4}. It was shown that in any given range of energy one can find $m=1$ states amidst others which are extended throughout the whole energy shell.

Since the $m=1$ states exist at all energies, it is interesting to determine the fraction of these states relative to the total number of states $N(E)$ as a function of energy. This question is important since in view of the Schnirelman theorem \cite{schni} stating that most eigenfunctions of a classically chaotic (ergodic) system are equidistributed over the energy shell as $\hbar\rightarrow 0$. In other words, we have
\begin{equation}
\lim_{E \rightarrow \infty} \frac{N_{m=1}(E)}{N(E)} = 0,
\label{lim}
\end{equation}
where $N_{m=1}(E)$ is the number of $m=1$ states up to some energy $E$. Neither the theorem nor Eq. (\ref{lim}) says anything about how the limit is approached. Moreover, the Schnirelman theorem does not consider the possible existence of parabolic fixed points. These occur in our billiard; they are the bouncing ball orbits, or the continuous set of all horizontal trajectories ($P_y=0$) which do not hit the boundaries. Quantum-mechanically, there are no $P_y=0$ states but the eigenstates of the flat channel with minimum value of $P_y$ are the $m=1$ states. Thus we sometimes refer to the $m=1$ states as ``bouncing ball states''. For $\epsilon\ll 1$ one can use the unperturbed spectra given by Eq. (\ref{e0}), to get a good estimate for $N_{m=1}(E)$ and $N(E)$. For large $N_{max}$ and $M_{max}$ the relation (\ref{e0}) represents, ignoring the $k^2$ term, half of an ellipse in the $n-m$ plane (recall that $m\in [1,M_{max}]$ and $n\in [-N_{max},N_{max}]$). Then for large $E$, the number $N(E)$ equals the area of the half ellipse, and the number $N_{m=1}(E)$ of bouncing ball states equals twice the size or its minor axis:
\begin{equation}
N(E) = \left( \frac{2m_e}{\hbar^2} \right) \frac{\sigma}{4\pi} E,
\label{n}
\end{equation}
\begin{equation}
N_{m=1} = \left( \frac{2m_e}{\hbar^2} \right)^{1/2} \frac{\sigma}{\pi} \frac{\sqrt{E}}{D},
\label{nm}
\end{equation}
therefore,
\begin{equation}
\frac{N_{m=1}(E)}{N(E)} = \left( \frac{\hbar^2}{2m_e} \right)^{1/2} \frac{4B}{\sigma \sqrt{E}},
\label{n/n}
\end{equation}
where $\sigma=BD$ is the area of one period of the billiard.

We see that Eq. (\ref{n}) is precisely the first term in the Weyl series for the integrated density of states (see for example \cite{stockmann}, Sect. 7). Equation (\ref{n/n}) indicates that i) for fixed $E$ and $\sigma$ the portion of $m=1$ states, relative to the total number, is larger the narrower the billiard is and ii) the convergence of $N_{m=1}(E)/N(E)$ to zero is rather slow. These estimates are in excellent agreement \cite{gala4} with detailed numerical calculations for the case $N_T=1$ and $\epsilon\ll 1$ (e.g., $\epsilon=0.06$).

For the case of a modulated boundary composed of many harmonics, the expression for $N_{m=1}$ is still valid. However, $N(E)$ is not simply given by (\ref{n}) since now the perimeter $\gamma$ of the boundary, as well as its curvature, contribute significantly to $N(E)$ as \cite{bh},
\begin{eqnarray}
N(E) & = & \left( \frac{2m_e}{\hbar^2} \right) \frac{\sigma}{4\pi} E - \frac{\gamma}{4\pi} \left( \frac{2m_e}{\hbar^2} \right)^{1/2} \sqrt{E} \nonumber \\ 
& \, & - \left( \frac{2m_e}{\hbar^2} \right)^{1/2} \frac{\sqrt{E}}{\pi} \sum_{r=1}^\infty \frac{(-1)^r}{(r-1/2)E^r} C_{2r+1}.
\label{weyl}
\end{eqnarray}

The coefficients $C_{2r+1}$, $r\geq 1$, in the sum of the Weyl series (\ref{weyl}) depend on the curvature (and its derivative) of the modulated boundary (see Table 3, p. 545 of \cite{bh}). As is shown in \cite{bh}, the higher the index $r$ of the Weyl coefficient $C_{2r+1}$, the more complex the expansions for $C_r$ are, involving higher and higher powers of the curvature and its derivatives. Clearly, the larger the number of harmonics $N_T$ is the more important the perimeter and curvature terms become. 

\section{Localization Lengths of Eigenstates versus Profile's Complexity}

A detailed analysis of the eigenstates for the billiard studied here, with rippled or rough boundaries, shows that there are three distinct types of eigenstates, namely, weakly, moderately, and strongly perturbed. To get a panorama of how these different types of states appear as a function of the energy, one may employ one of various quantities that give a measure of the effective number of unperturbed states that form the eigenstates of the perturbed system. One such measure is the ``entropy localization length'' \cite{gala3} defined by
$l_H =  \exp {\left[-\left( \aleph - \aleph_{GOE}\right)
\right]} \approx 2.08 \exp {(-\aleph)}$. Here $\aleph$ is the Shannon entropy 
of an eigenstate in a given basis, 
${\aleph}=\sum^{L_{max}}_{l=1} \mid C^{\alpha}_l\mid^2 \,\ln \mid C^{\alpha}_l\mid^2$, and $\aleph_{GOE}$ is the entropy of a completely chaotic state which is characterized by gaussian fluctuations (for $L_{max}\rightarrow \infty $) of all components $C_l^\alpha$ with the same variance $<\mid C^{\alpha}_l\mid^2>=1/L_{max}$. Fig. 3 shows $l_H$ as a function of the level number $\alpha$ [equivalently, as a function of the integral density of states $N(E)$] for the cases of $N_T= 2, 4, 6,$ and $100$ harmonics, respectively. In all cases we see that with the level number $\alpha$, $l_H$ increases on average. This is expected since with the increase of energy the eigenstates are expected to be more ergodic. Also, these figures demonstrate two other features which are only proper of quasi 1D and 2D billiards with small amplitude of rippleness or roughness on the boundaries: i) the increase on the number of states with minimal $l_H$ as the complexity of the profile increases and ii) the strong and rapid fluctuations of $l_H$ as a function of $\alpha$. The first feature is easy to understand if we recall that the expression (\ref{nm}) for $N_{m=1}$ is valid regardless of the complexity of the boundary while $N(E)$ does depend on the complexity. In fact, as $N_T$ increases the second term in (\ref{weyl}) becomes more influential and hence the ratio $N_{m=1}/N(E)$ increases with $N_T$. 

\begin{figure}
\centerline{\epsfig{figure=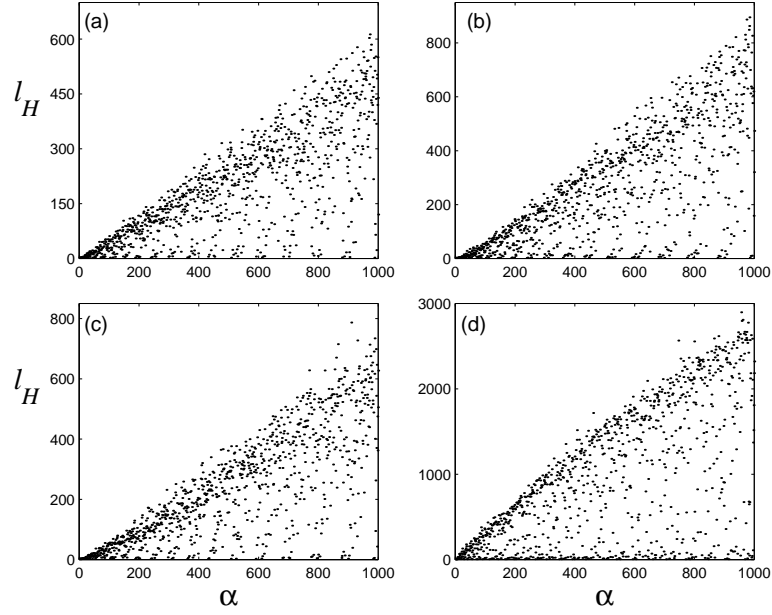,width=4in}}
\caption{Localization measure $l_{H}$ for the first 1000 ordered in energy eigenfunctions for (a) $N_T=2$, (b) $N_T=4$, (c) $N_T=6$, and (d) $N_T=100$.}
\label{fig3}
\end{figure}

For any strongly localized state it is possible to determine the pair $(n,m)$ of the unperturbed state that determines the main component of such localized state. Fig. 4 shows two sets of strongly localized states that have been identified with the unperturbed states $(n,m)=(-18,1)$ and $(n=-18,m=2)$. The first row of states corresponds to the $m=1$ unperturbed states and the second row to the $m=2$ states. Note that the level number $\alpha$ corresponding to these states shifts down gradually as $N_T$ increases, in agreement with the observation made above that the ratio $N_{m=1}/N(E)$ increases as $N_T$ increases. A quick glance of figures 4(a1-c1) also gives the impression that all $m=1$ states are qualitatively the same. However, some striking differences between them come to light when viewed in the configuration representation. Fig. 5 shows the same sequence of states as of Fig. 4 but in the configuration representation. In this figure we see the one maximum and two maxima of the wave functions which are characteristic of the $m=1$ and $m=2$ states, however note that the wave function for the case $N_T= 100$ has become drastically asymmetric with respect to $y=d/2$; it is ``repelled'' by the rough boundary. We shall analyze this unexpected phenomenon in the next section.

\begin{figure}
\centerline{\epsfig{figure=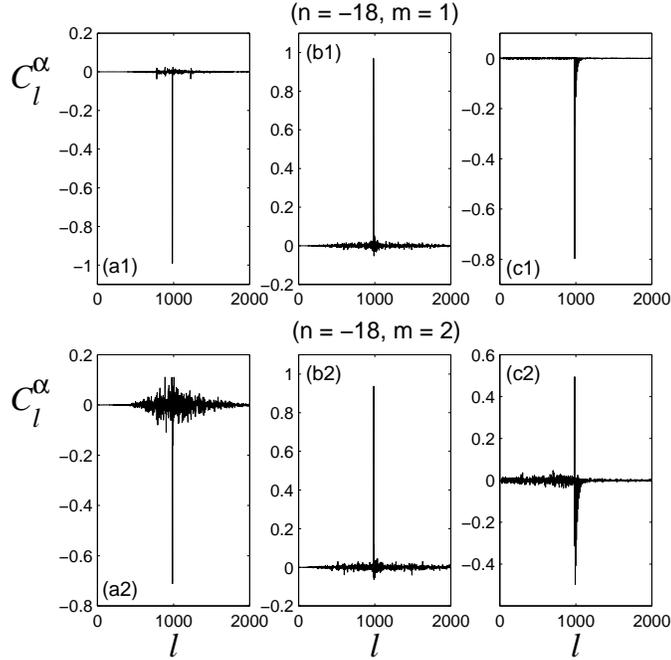,width=3.5in}}
\caption{Localized eigenfunctions in the energy representation for (a1) $N_T=2$ ($\alpha = 987$), (b1) $N_T=6$ ($\alpha = 983$), (c1) $N_T=100$ ($\alpha = 391$), (a2) $N_T=2$ ($\alpha = 990$), (b2) $N_T=6$ ($\alpha = 986$), and (c2) $N_T=100$ ($\alpha = 396$).}
\label{fig4}
\end{figure}

\begin{figure}
\begin{center}
{\bf \Large FIGURE 5}\\

{\small (available in gif format)}
\end{center}
\caption{Eigenfunctions of Fig. 4 in the configuration representation $\mid \psi^{\alpha}(x,y;k) \mid ^2$; (a1) $N_T=2$ ($\alpha = 987$), (b1) $N_T=6$ ($\alpha = 983$), (c1) $N_T=100$ ($\alpha = 391$), (a2) $N_T=2$ ($\alpha = 990$), (b2) $N_T=6$ ($\alpha = 986$), and (c2) $N_T=100$ ($\alpha = 396$).}
\label{fig5}
\end{figure}

The second feature revealed by the plots of $l_H$ as a function of $\alpha$ (Fig. 3) namely, the strong and rapid fluctuations, reflects the coexistence of localized, sparse and extended states, characterized respectively, by small, medium, and large values of $l_H$. It was argued in the previous section that the number of unperturbed states conforming a given exact state depends on the value of the transversal mode $m$ of the original unperturbed state. Since neighboring states in the energy-ordered unperturbed spectra generally have completely different values of the transversal number $m$, say, $n=N_{max}, m=1$ and $n=1, m=M_{max}$, it follows that the perturbed spectra should show rapid and strong fluctuations of $l_H$ as a function of energy or $\alpha$. This is expected to be independent of the complexity of the boundary. For the case $N_T=100$, Fig. 6. shows a typical sequence of consecutive levels illustrating that if we arbitrarily pick up two neighboring states, knowledge of the structure of one does not allow us to predict the structure of the next.

\begin{figure}
\centerline{\epsfig{figure=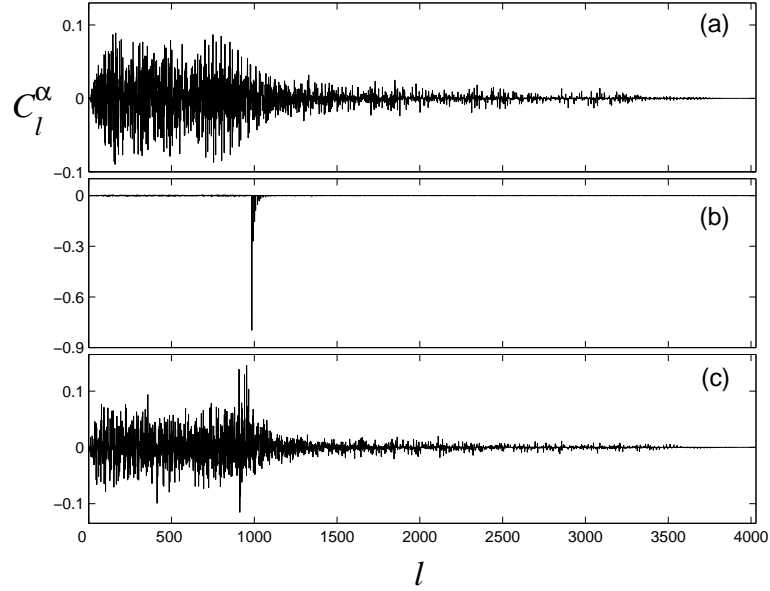,width=4in}}
\caption{Typical consecutive eigenfunctions in the energy representation for the case $N_T=100$. (a) $\alpha = 390$, (b) $\alpha = 391$, and (c) $\alpha = 392$.}
\label{fig6}
\end{figure}

\begin{figure}
\centerline{\epsfig{figure=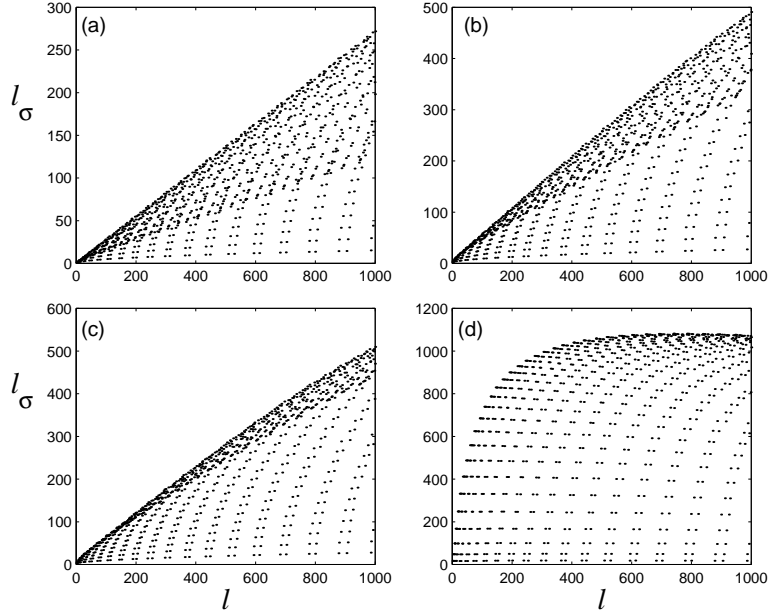,width=4in}}
\caption{Localization measure $l_{\sigma}$ for the first 1000 ordered in energy unperturbed states expanded in the basis of $\hat{H}$ for (a) $N_T=2$, (b) $N_T=4$, (c) $N_T=6$, and (d) $N_T=100$.}
\label{fig7}
\end{figure}

Additional information can be obtained by considering the reverse process, namely, the expansion of an unperturbed state in the basis of the perturbed system. In particular, we can study the ``width'' $l_{\sigma}$ or mean square root of an unperturbed state $\phi_l^0$ defined as $l_\sigma= \{ \, \sum^{L_{max}}_{\alpha=1} \mid C^{\alpha}_l\mid^2\,\left[\alpha-n_c(l)\right]^2 \}^{1/2}$, where $n_c = \sum_\alpha \alpha \, \mid C^{\alpha}_l\mid^2$ gives the centroid of an unperturbed state in the perturbed basis. Fig. 7 reports the plot of $l_{\sigma}$ as a function of $l$ for $N_T= 2, 4, 6$, and $100$. One can see a clear regularity in these plots. The lowest row corresponds to unperturbed states with $m=1$ and $m=2$. The next row correspond to unperturbed states with $m=2$ and $m=3$, and so on, see \cite{gala3} for details. We see that as  $N_T$ increases the pattern of pairs persists for higher and higher values of the quantum number $m$. This fact clearly reflects a regular character of the states with low values of $m$ independently of how large the energy is. Note that the energy is not the unique semiclassical parameter, in fact we have two such parameters, $m$ and $n$.

\section{Structure of Energy-Localized States for Disordered Profiles}

Here we concentrate on the nature of eigenstates for the case of a rough boundary, $N_T= 100$. In Fig. 8(a1-b2) we present a typical pair of extremely energy-localized states together with their respective enlargements. Both are $m=1$ states but with different values of $n$, namely $(n,m)=(-18, 1)$, and $(-28,1)$. Close inspection of Figs. 8(a2-b2) shows that, unlike the case of profiles with small number of harmonics $N_T$, these states reveal other components of appreciable magnitude. A decoding of these components  shows that the second most important component of each of these states corresponds to $m=2$, the next one to $m=3$ and so on, all with the same value of $n$. That is, there is no mixing between different values of $n$. Moreover, as Figs. 8(a2-b2) suggest, their amplitude decay exponentially. We have found that, indeed, this is the case and hence one can write [see Eq. (3)],
\begin{equation}
C_l^{\alpha}(k) = C_{mn}^{\alpha}(k) \cong S\exp[-\beta(m-1)] \, \delta_{n,n_{\alpha}}
\label{cal}
\end{equation}
where $n_{\alpha}$ corresponds to the energy-localized state. The parameters $S=0.8$ and $\beta =0.53$ are obtained by fitting the data to the exponential dependence. It is important to remark that for a fixed $N_T$ all energy-localized states give the same values of $S$ and $\beta$. Substituting Eq. (\ref{cal}) into Eq. (3) gives
\begin{eqnarray}
\psi_{loc}^{\alpha}(u,v;k) & = & C S \pi^{-1/2} g^{-1/4} \exp\left[{i(k+n_\alpha)u}\right] \nonumber \\ 
& \, & \times \, \sum_{m=1}^{\infty} \sin \left( \frac{m \pi v}{d} \right) \exp\left[{-\beta(m-1)}\right], 
\label{eq16}
\end{eqnarray}
where $C$ arises to satisfy the orthonormality condition in curvilinear coordinates,
\begin{equation}
\int_0^{2\pi} \int_0^1 du dv \sqrt{g} \psi_{loc}^{\alpha \ *} \psi_{loc}^\alpha = 1. 
\end{equation}

\begin{figure}[htb]
\centerline{\epsfig{figure=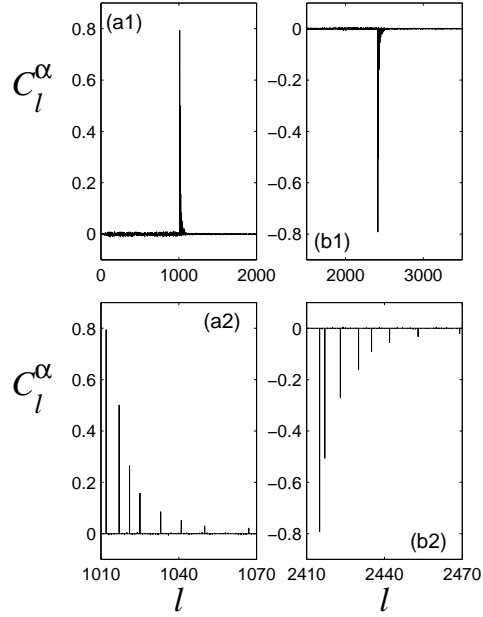,width=2.5in}}
\caption{Localized eigenfunctions in the basis representation (upper part) 
and an enlargement of the main components of them (lower part) for $N_T=100$; 
(a) $\alpha = 398$ and (b) $\alpha = 968$. The main component corresponds to the unperturbed state characterized by (a) $(n,m)=(18,1)$ and (b) $(n,m)=(-28,1)$. The components shown in (a2-b2) correspond to unperturbed states with $m=$1, 2, 3, 4, 5, 6, 7, and 8, respectively (from left to right).}
\label{fig8}
\end{figure}

It follows that 
\begin{eqnarray}
\mid \psi_{loc}^{\alpha}(u,v;k) \mid ^2 & = & \pi^{-1} g^{-1/2} \left( 
\sum_{m=1}^{\infty} \exp\left[{-2\beta(m-1)}\right] \right)^{-1} \nonumber \\
& \, & \times \, \left[ \sum_{m=1}^{\infty} \sin \left( \frac{m \pi v}{d} \right)  \exp\left[{-\beta(m-1)}\right] 
\right]^2 
\end{eqnarray}
and
\begin{eqnarray}
\mid \psi_{loc}^{\alpha}(y) \mid ^2 & = & \pi^{-1} \left( 
\sum_{m=1}^{\infty} \exp\left[{-2\beta(m-1)}\right] \right)^{-1} \nonumber \\
& \, & \times \, \left[ \sum_{m=1}^{\infty} \sin \left( \frac{m \pi v}{d} \right) \exp\left[{-\beta(m-1)}\right] 
\right]^2 \nonumber \\ 
& = & \frac{1-\exp(2\beta)}{4\pi} \frac{\sin^2(\pi y/d)}{\cos(\pi y/d)-\cosh(\beta)}.
\label{philoc}
\end{eqnarray}

This latter expression gives the average shape of the localized eigenstates in configuration representation, projected onto the $y-\mid \psi^\alpha \mid ^2$ plane. Also note that $\psi_{loc}^{\alpha}$ does not depend on the parameter $S$, see Eq. (14). To obtain (\ref{philoc}) we have made the approximation $v \approx y$ and $g \approx 1$ since $\epsilon=0.06\ll 1$. In Fig. 9 we plot Eq. (18) together with the numerical data for the state $\alpha=391$, see also Figs. 4(c1) and 5(c1). The excellent agreement indicates that for extremely localized states with $m=1$ there is no need to diagonalize the whole matrix since there is no mixing between different values of $n$. In order to obtain a good approximation of these eigenstates one only needs to diagonalize the block of size ($M_{max} \times M_{max}$) corresponding to the quantum number $n_{\alpha}$, which is a good quantum number in the presence of perturbation. In the inset of Fig. 9 we show global shapes of the $m=1$ states for different values of $N_T$. We have checked that these curves fit the $m=1$ states very well. Since the parameters $S$ and $\beta$ are the same for all $m=1$ states for a fixed $N_T$ one can find $S$ and $\beta$ for some low energy state (so that the matrix to be diagonalized is small) and with this one can find the global shape of {\it all} $m=1$ states. Conversely, knowing the shape of a $m=1$ state within any energy range, one may assess by looking at the amount of repulsion, the degree of complexity, presumably unknown, of the boundaries. 

\begin{figure}
\centerline{\epsfig{figure=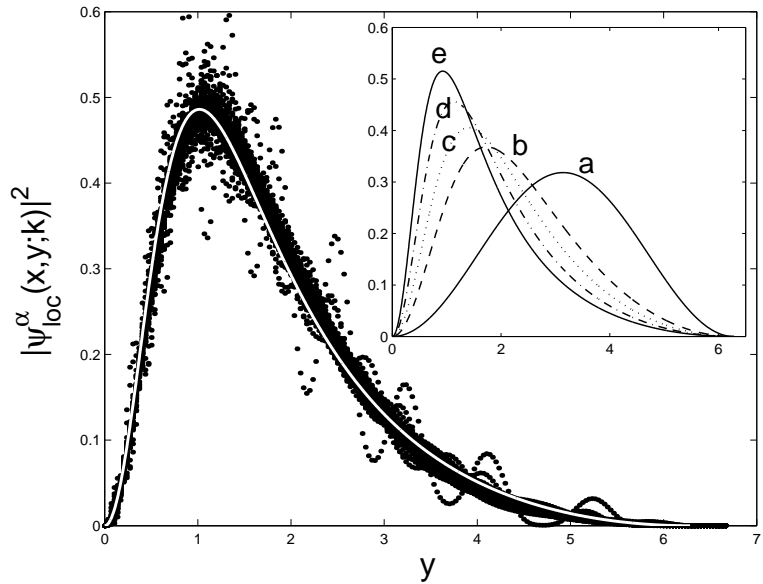,width=4in}}
\caption{Density $\mid \psi^{\alpha}(x,y;k) \mid ^2$ (points) for $\alpha=391$ and prediction (\ref{philoc}) for strongly localized eigenstates for $N_T = 100$: $\mid \psi_{loc}^{\alpha}(y) \mid ^2$ (white line) [with $\beta=0.53$]. Inset: $\mid \psi_{loc}^{\alpha}(y) \mid ^2$, Eq. (\ref{philoc}), for (a) $N_T=1$, $\beta=0$, (b) $N_T=25$, $\beta=1$, (c) $N_T=50$, $\beta=1.3$, (d) $N_T=75$, $\beta=1.67$, and (e) $N_T\rightarrow \infty$, $\beta=2.08$.}
\label{fig9}
\end{figure}

\begin{figure}
\begin{center}
{\bf \Large FIGURE 10}\\

{\small (available in gif format)}
\end{center}
\caption{Localized eigenfunctions for $N_T=100$ in the basis and configuration representation; (a) $\alpha = 976$ and (b) $\alpha = 990$.}
\label{fig10}
\end{figure}

Thus, the states most localized in energy, correspond to the smallest values of $m$ of the originally unperturbed system. One can expect the same type of behavior, namely, the repulsion of  the wave function away from the rough boundary for $m=2$, see Fig. 10. However the main components of the wave function belong to different combinations of $n$ and $m$. Nonetheless, their amplitudes decay roughly exponentially. One may treat the repulsion effect as the inability of the wave function to resolve the fine details of the rough profile and hence to satisfy the Dirichlet boundary conditions. As the value of $m$ of the originally unperturbed state becomes larger, details of the profile can be resolved better and consequently the probability density should become appreciable in the neighborhood of the rough boundary. In the next section we shall see that this is indeed what happens.

\section{Structure of the Energy-Extended States: Transition from Smooth to Rough Boundaries}

Here we study the nature of those eigenstates of the billiard with large values of localization lengths. These are selected from plots of entropy localization lengths, like those of Fig. 3, for a given number of harmonics. In Fig. 11 we present a typical set of this type of states in energy representation for
$N_T=1$, 50, and 100. All of these states were chosen to be of approximately the same energy, namely $E^{\alpha}$ around the energy of the unperturbed level $l=980$. Note that their level number ${\alpha}$ decrease as $N_T$ increases because, as we showed in Section 2, the level number $N(E)=\alpha$ grows slower as a function of energy as the complexity of the profile increases. For smooth profiles with $N_T\sim 1$ the amplitude of its components decay roughly symmetrically around $l=980$. However, as $N_T$ increases the states spread more towards the high-lying unperturbed states with not a single predominant component.

\begin{figure}
\begin{center}
{\bf \Large FIGURE 11}\\

{\small (available in gif format)}
\end{center}
\caption{Extended eigenfunctions $C^\alpha_l$ in the basis representation (left) and in the configuration representation $|\psi^\alpha(x,y;k)|^2$ (right); $\alpha = 986$ for $N_T=1$, $\alpha = 582$ for $N_T=50$, $\alpha = 407$ for $N_T=100$.}
\label{fig11}
\end{figure}

The strong asymmetry can be understood as follows. As the perturbation increases, more and more components are needed to construct the exact states of the modulated billiard. Since the unperturbed levels are bounded from below, higher-lying unperturbed states are needed to form an exact state. Another argument rests on the classical-quantum correspondence of the structure of eigenstates \cite{gala2,gala3,gala4,CCC}. The structure of an eigenstate $\alpha$ is given by {\it the average} $W(E^0 \mid E^\alpha)$ {\it over a number of states around the state $\alpha$}. In \cite{gala3} it is shown how to construct the {\it{classical}} counterpart of the eigenstates (CCSE) for non-integrable billiards. Briefly, in analogy with the quantum case where we project the eigenstate of the full Hamiltonian $\hat{H}$ on the basis of $\hat{H}^0$, classically we project the dynamics of $H$ onto $H^0$. Operationally, this is done by substituting the solutions of $H$ into $H^0$. Specifically, we write the Hamiltonian in the curvilinear coordinates $(u,v)$. Thus the classical Hamiltonian for our billiard can be written as $H=H^0+\epsilon V$, where,
\begin{equation}
H^0= \frac{1}{2m_e}(P_u^2+P_v^2)
\label{h0}
\end{equation}
and
\begin{equation}
V=-\frac{1}{2m_e}\left[ \frac{2\xi+\epsilon(\xi^2 +v^2 \xi_u^2)}{(1+\epsilon \xi)^2}P_v^2 +\frac{2v\xi_u}{1+\epsilon \xi}P_v P_u \right]
\label{v}
\end{equation}

By substituting the solutions of $H$ into $H^0$, $H^0$ becomes a function of time, varying within an upper $H^0_{up}$ and a lower $H^0_{low}$ limit. The energy range spanned by the projection of the dynamics of the full Hamiltonian onto $H^0$ is $\Delta E=H^0_{up}-H^0_{low}$, referred to as the width of the ``energy shell''. Since the dynamics of $H$ is fully chaotic, it is enough to substitute a single trajectory of $H$ into $H^0$ and to let it run for a sufficiently long time to span the whole available phase space. From this, one obtains the distribution of energy values which is the CCSE in energy space \cite{gala3}.

In Fig. 12 we show the structure of some eigenstates and their classical counterpart for $N_T=2$ and $10$. Quantum oscillations around the classical curve are found to be quite strong because the average generally includes extended and localized or sparse states \cite{gala3}. Moreover, in average, the range of the energy shell and the shape of the CCSE correspond quite well to the global shape of eigenstates.

Below we analyze the particular peak structure of these curves but before let us discuss the dynamical origin of the observed increase of $H^0_{up}$ as a function of $N_T$. We first note that the new canonical momenta are given by
\begin{equation}
P_u=P_x +\epsilon \frac{yP_y\xi_x}{(1+\epsilon \xi)}
\label{pu}
\end{equation}
and 
\begin{equation}
P_v=(1+\epsilon \xi)P_y,
\label{pv}
\end{equation}
where $P_x$ and $P_y$ are both within the range $[-\sqrt{2m_eE},\sqrt{2m_eE}]$, with $E$ being the energy of the total Hamiltonian. From Eqs. (\ref{pu}) and (\ref{pv}) we see that $P_v$ is of the same order as $P_y$ but $P_u$ can be much larger than $P_x$ since the derivative $\xi_x=-\sum_N^{N_T} N A_n \sin(Nx)$ can be very large, roughly proportional to $N_T$ for large $N_T$. 

\begin{figure}
\centerline{\epsfig{figure=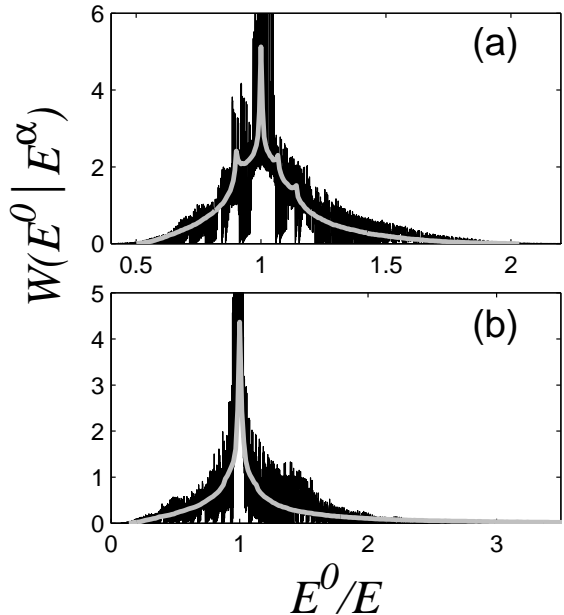,width=3in}}
\caption{Structure $W(E^0_l \mid E^\alpha)$ of eigenstates in the energy representation (broken curve) and the CCSE (thick smooth curve), as a function of the normalized energy $E^0/E^\alpha$. The average is taken over the interval (a) $2200< \alpha <2400$ for $N_T=2$, and (b) $2200< \alpha <2320$ for $N_T=10$.}
\label{fig12}
\end{figure}

The upper limit $H^0_{up}$ of the energy shell then ought to grow roughly proportional to $N_T^2$. Inspection of plots of Fig. 13 shows that indeed this is the case. For example, for $N_T=50$ and $N_T=25$, the ratio $H^0_{up,50}/H^0_{up,25}=(50/25)^2=4$, which agrees well with the limits found numerically. 

\begin{figure}
\centerline{\epsfig{figure=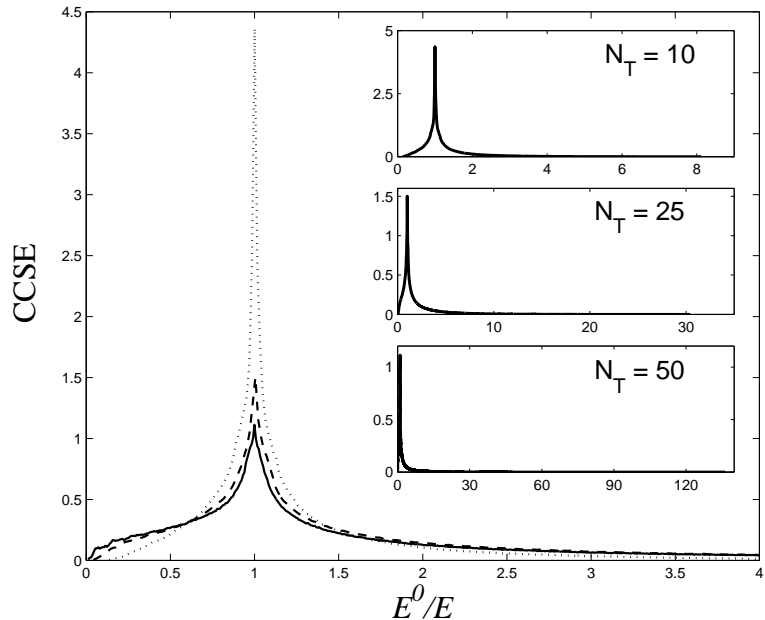,width=4in}}
\caption{CCSE for $N_T=10$, $25$, and $50$.}
\label{fig13}
\end{figure}

Now let us study the form of eigenstates as a function of $N_T$. For the case $N_T=2$ in Fig. 12(a) we see that the CCSE has four peaks in a rough agreement with the global shape of eigenstates. For $N_T=1$ in \cite{gala3} it was shown that both classical and quantum structures have three peaks, a central one and one on each side of the central peak. It was found that the main contribution to the central peak comes from grazing trajectories (bouncing balls), while the right (left) peak is due to trajectories dwelling in the neighborhood of the period one stable (unstable) fixed point. The profile $\xi(u)=A_1\cos(u) + A_2\cos(2u)$ for the case $N_T=2$ gives rise to three periodic orbits of period one; two stable ones at $(P_u,u)=(0,0)$ and $(0,\pi)$ and one unstable at $(P_u,u)=(0,\cos^{-1}(-A_1/4A_2))$. At these points, the values of the profile function are: $\xi(0)=A_1+A_2, \xi(\pi)=-A_1 + A_2$, and $\xi(\cos^{-1}(-A_1/4A_2))=-(A_1^2/8A_2)-A_2$. Substituting $P_u=0$ and $\xi_u=0$ into (19-20) one obtains $P_v^2=2m_eH(1+\epsilon \xi)$ and $H^0=(1+\epsilon \xi)^2H$, where $\xi$ stands for $\xi(u)$ evaluated at the fixed period one points. Inserting the randomly generated values of $A_1$ and $A_2$ used in the calculations, we obtain $H^0=0.899$, $1.0681$, and $1.1409$, corresponding to the periodic points at $u=0$, $\pi$, and $\cos^{-1}(-A_1/4A_2)$, respectively. The two peaks on the right of the highest peak of the CCSE of Fig. 12(a) correspond to the stable period one orbits and the left peak to the unstable period one orbit.

Clearly, the number and location in energy of the peaks depend on the coefficients $A_N$. Then by looking at the structure of the eigenstates, one can infer about the complexity of the profile. As the number of harmonics determining the profile increases, $\Delta E$ grows proportional to $N_T^2$ and all the peaks produced by periodic orbits of period one, which are located very close to the central peak at $H^0=H$, seem to disappear in favor of a wider peak around $H^0=H$.

Thus, the CCSE allows us to understand characteristic features of the eigenstates in terms of the underlying classical dynamics.

\section{Energy Band-Structure: Localized versus Extended States}

We address here the question of how the energy-localized and energy-extended states show up in the energy band structure of the billiard as the complexity of the profile increases from smooth to rough.

We have seen that the energy localized states $\psi_{loc}^\alpha(u,v;k)$ are well described by Eq. (\ref{eq16}). Since they are very close to unperturbed states, their normalized energy $E^\alpha/E^*$ is given approximately by 
\begin{equation}
E^\alpha/E^* \approx \frac{d^2}{\pi^2}\left [(k + n_\alpha)^2 + \frac{\pi^2}{d^2} \right ],
\end{equation}
and the group velocity (the slope of the bands) is
\begin{equation}
V_g=\frac{\partial (E^\alpha/E^*)}{\partial k} \simeq \frac{d^2}{\pi^2} (k + n_\alpha) = 8(k+n_\alpha),
\end{equation}
where $E^\alpha\cong E^0_{n,m}(k)$ and $E^*=(\hbar^2/2m_e)(\pi/d)^2$.

In the last equation we substituted $d=2 \pi$, the value used throughout this paper. The group velocity $V_g$ as given by Eq. (24) should be very close to the exact one regardless of the value of $N_T$. To check our expectation let us analyze Fig. 14 which shows a segment of the energy-band structure for the case $N_T= 100$ in the range $0.05 < k < 0.15$, and $1280 < E^\alpha/E^* < 1370$. Note that there are mostly only two types of bands, flat bands and straight sloped bands, everywhere except in the neighborhood of avoided crossings. In this figure we indicate the position of some representative eigenstates at $k=0.1$ which are shown in their energy and configuration representation in Fig. 15. Inspection of these two figures reveal that: i) eigenstates lying on horizontal (inclined) bands correspond to energy-extended (localized) states, and ii) the slopes computed directly from the graphs agree very well with the group velocity given by Eq. (24). Therefore, the inclined bands are actually parabolic bands but because the range in $k$ is small, they appear in this figure as straight sloped lines. The agreement is quantitative, for example, the slope of the bands at $k=0.1$ for states 391 and 408 is $\pm 245.5$. The use of Eq. (24) with $k=0.1$ gives $n_\alpha =18.09$, with the integer part 18 being exactly what is found by identifying the $(n,m)$ pair of the unperturbed state that coincides with the main component of the energy-localized state ($\alpha=391$ or 408). Such a good agreement is surprising for the case of the state $\alpha=391$ since it lies near an avoided crossing, where one may anticipate strong mixing of many unperturbed components. We have examined the structure of the states in the neighborhood of the avoided crossing as we vary $k$ (see inset of Fig. 14) and observed a gradual exchange of identities between states. Specifically, for $k=0.1022$ the state $\alpha=390$ (391) is extended (localized) in energy, and as $k$ increases the state $\alpha=391$ acquires more and more components till at $k=0.1034$ it becomes extended throughout the whole energy shell, looking very much like the state $\alpha=390$ at $k=0.102$. Thus, the states retain their character (extended or localized) throughout the Brillouin zone except in a small neighborhood of the avoided crossings.

\begin{figure}[htb]
\centerline{\epsfig{figure=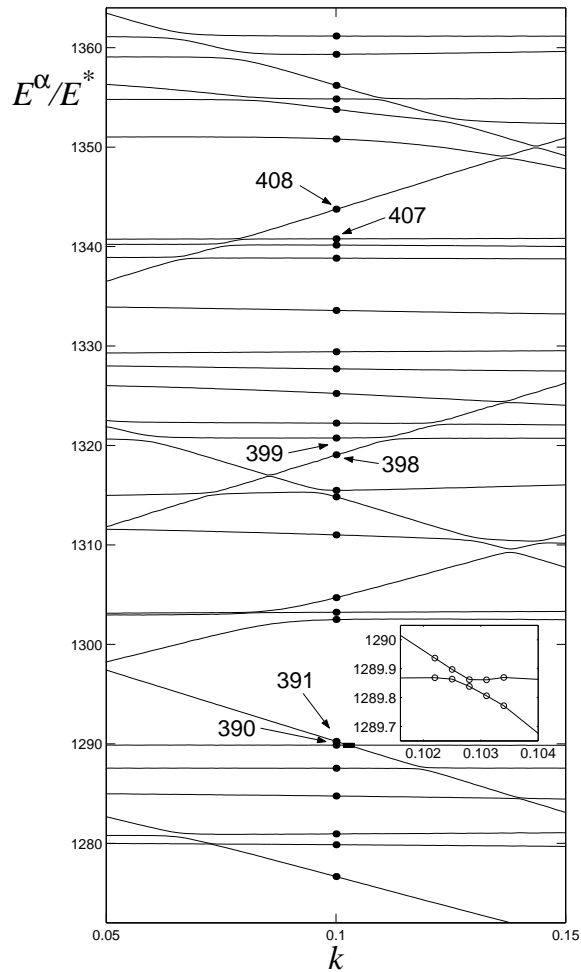,width=3in}}
\caption{Energy band structure for $N_T=100$, and $(a/2\pi,d/2\pi)=(0.06,1.0)$. Inset: enlarged avoided crossing marked as a black rectangle at the right of the states $\alpha = 390$ and 391.}
\label{fig14}
\end{figure}

\begin{figure}
\begin{center}
{\bf \Large FIGURE 15}\\

{\small (available in gif format)}
\end{center}
\caption{The eigenstates $\alpha=390$, 391, 407, and 408, marked in Fig. 14, in the basis representation (left) and in the configuration representation (right).}
\label{fig15}
\end{figure}

\begin{figure}
\centerline{\epsfig{figure=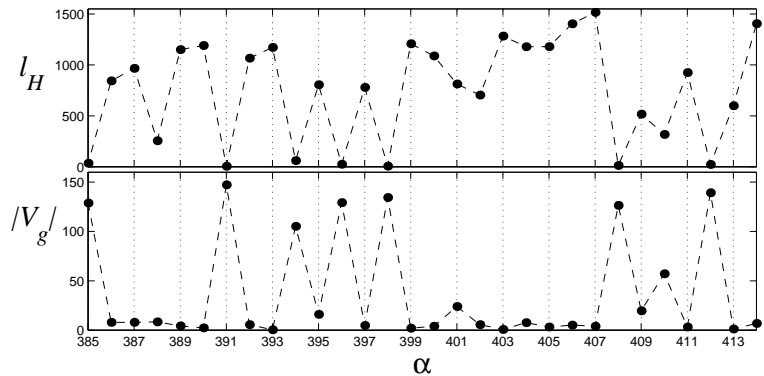,width=4in}}
\caption{Entropy localization length $l_H$ and group velocity $|V_g|$ for the eigenstates in the range of energy of Fig. 14, $\alpha=[385,414]$ ($N_T=100$, $k=0.1$).}
\label{fig16}
\end{figure}

It should be stressed that for any high energy one can find eigenstates with a very different structure. Note that the energy-extended states of Fig. 15 are strongly localized in the coordinate $x$ within one period of the billiard. We found that in the case of rough profiles, about a third of energy-extended states become localized in configuration representation. This localization effect is in agreement with the structure of the corresponding energy band spectra: since these states have $V_g\sim 0$ their wave functions are not allowed to propagate through the billiard. Also from Fig. 11 one can clearly see that the transition from smooth to rough boundary in the billiard leads to a transition from extended to localized in configuration representation of the energy-extended states.

Since states located on flat (parabolic) bands correspond to energy-extended (-localized) states, their localization lengths are expected to be roughly proportional to $\mid V_g \mid ^{-1} $. A detailed comparison shows that while this inverse proportionality is qualitatively correct, the group velocity does not distinguish between fully and partially energy-extended states, they both have $\mid V_g \mid $ close to zero. This is illustrated in Fig. 16. Nevertheless, it is clear that the group velocity gives useful information about the nature of a given state and, besides, it is the band structure which is usually known, and not the wavefunction itself. Thus, a graph of $V_g$ as a function of $\alpha$, for an arbitrary value of $k$ within the Brillouin zone can be very helpful in understanding the global structure of the energy band and the type of corresponding eigenstates, localized or extended in energy space.

Fig. 17 shows $l_H$ and $\mid V_g \mid$ as a function of $\alpha$ for $N_T=1$, 25, 50, 75, and 100. We consider first the case $N_T=100$, noting the dominance of states with zero or near zero group velocities (equivalently, maximal values of $l_H$). This is consistent with Fig. 14 but Fig. 17 also strongly suggests that both types of bands (flat and parabolic) and, correspondingly, the both types of eigenstates (extended and localized in energy) appear at all energy ranges, with the exception of the lowest levels where the perturbative regime holds. Further, the sequence of graphs in Fig. 17 reveal that the distribution of slopes or entropy localization lengths changes gradually as the complexity of the profile changes. Ona can compare the plots of $|V_g|$ as a function of $\alpha$ of Fig. 17 for $N_T=1$ and $100$. The former shows the no-proliferation of flat bands, in fact, the values of $\mid V_g \mid$ appear evenly distributed: there are more intermediate than maximal or minimal values of $\mid V_g \mid$. It would imply fluctuations of the bands yielding an spaghetti-like structure. This is indeed the case which is qualitatively different from that of the rough profile discussed above. The rest plots of Fig. 17 imply that there is a gradual transition in the energy band structure as $N_T$ increases and this in turn means that as the underlying classical dynamics goes from chaotic to random the eigenstates with intermediate localization lengths tend to disappear in favor of either extremely extended (in energy) or extremely localized (in energy) states. Hence, knowledge of the group velocity gives information not only about the type of energy band spectra but also about the underlying classical dynamics, in its transition from chaos to disorder.

\begin{figure}[t]
\centerline{\epsfig{figure=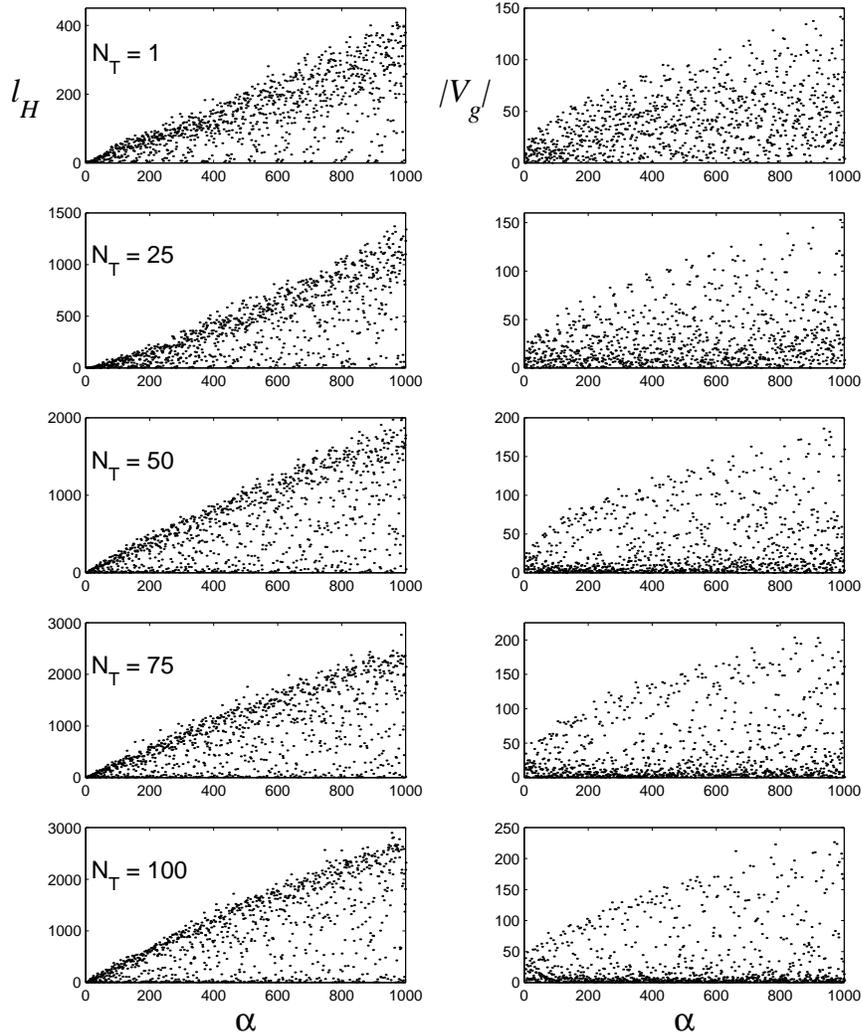,width=4.5in}}
\caption{Entropy localization length $l_H$ (left) and group velocity $|V_g|$ (right) for the first 1000 eigenstates for $N_T=1$, 25, 50, 75, and 100; $k=0.1$.}
\label{fig17}
\end{figure}

\section{Conclusions}

We have studied the transition from chaos to disorder in periodic billiards, focusing on the properties of eigenstates and energy band spectra. We did this by considering a prototype 2D billiard, defined by two parallel walls, one wall is flat and the other is defined by a sum of $N_T$ harmonics with randomly distributed amplitudes. The larger the sum the more complex the profile, and for a large $N_T$ the profile becomes rough (disordered). We remark that we considered the case where the root mean square of the amplitude $a$ is small compared to the the average width $d$ of the billiard, ($a/d \ll 1$).

We found that independent of the number of harmonics, and for any energy range, there are extremely localized states co-existing with sparse and extended states (in the unperturbed energy basis). These ``energy-localized states'' originate from unperturbed states with the smallest values of the transversal mode $m=1$, 2, 3. Being very similar to traveling waves, these states have their maximum support along one ($m=1$) or two ($m=2$) horizontal strips. Hence they are identified as bouncing ball states. As the complexity of the profile increases, these strips are gradually repelled towards the flat wall. Hence, knowledge of the amount of repulsion gives an indication of how smooth or complex the modulated profile is.

We showed that the number of the energy-localized states relative to the total number of states up to some energy $E$ is proportional to $E^{-1/2}$, hence the approach to ergodicity is rather slow. Further, for small $N_T$, these energy-localized states coexist in roughly equal numbers with sparse and extended states but as  the complexity of the profile increases the number of sparse states decreases in favor of ergodic states. We found that in the case of rough profiles, about a third of energy-extended states become localized within one period of the billiard in configuration space.

While analyzing the energy band structure, we showed that the ``dynamics'' of the levels for the case of {\it rough} boundaries is quite simple: they follow either horizontal or parabolic trajectories in $k$-space, except very near the avoided crossings. States lying on horizontal (parabolic) bands are energy-extended (localized) states. In contrast, for smooth boundaries $N_T\sim 1$, the bands form the typical spaghetti-like structure of chaotic systems. Thus, the energy band structure reflects clearly the degree of complexity of the boundaries and, in turn, the transition from chaos to disorder. It was also shown that the group velocity $V_g=dE/dk$ is a very useful quantity for understanding dynamical features of the system. Specifically, plots of $V_g$ as a function of $\alpha$, where $\alpha$ labels the energy-eigenstate of the perturbed Hamiltonian, show distinct patterns for smooth and rough boundaries. The utility lies in that it is only necessary to calculate $V_g$ as a function of $\alpha$ for any value of $k$, to predict the type of energy band structure and also the type of underlying classical dynamics (chaotic or disordered).

\vspace{0.5cm}
{\bf Acknowledgements}: We wish to acknowledge financial support from CONACyT (Mexico) Grant No. 34668-E, and from project VIEP II63G02, BUAP.

\end{document}